# Copolymerization of partly incompatible monomers: an insight from computer simulations.

Alexey A. Gavrilov[1], Alexander V. Chertovich

*Physics Department, Lomonosov Moscow State University*

In this work we used dissipative particle dynamics simulations to study the copolymerization process in the presence of spatial heterogeneities caused by incompatibility between polymerizing monomers. The polymer sequence details as well as the resulting system spatial structure in the case if phase segregation occurs during the chain growth can be predicted using the method. We performed the model verification with the available literature data on styrene-acrylic acid copolymerization in the bulk and a very good agreement between experimental and simulated data for both chain average composition and triad fractions was observed. Next, we studied the system properties for a model symmetric reaction process with the reactivity ratios $r_1 = r_2 = 0.5$ at different compositions and Flory-Huggins parameters $\chi$. We found that the system average copolymer "composition-feed" curve does not depend on the $\chi$-value, but there are significant changes in the copolymer sequences caused by the density fluctuations. Finally, we investigated the formation of gradient copolymers during living styrene-acrylic acid copolymerization at a highly asymmetrical feed composition.

## 1. Introduction.

During the last decades the controlled radical polymerization has become the main polymerization technique in the majority of polymer laboratories; it includes several schemes which are suitable for synthesizing polymer molecules of different architecture. The three most widely used methods are the Reversible Addition Fragmentation Chain Transfer (RAFT) polymerization, the Atom Transfer Radical Polymerizations (ATRP), and the Nitroxide Mediated Living Free Radical Polymerizations (NMP).[1] At the same time so-called Polymerization-Induced Phase Separation (PIPS) has already become one of the most promising technique to obtain microstructured polymer materials;[2] it has been recently used to prepare crosslinked glass-like microheterogeneous thermoset resins.[3] Arresting phase separation at the nanoscale during the growth of immiscible polymers is a simple and effective route to obtain nanostructured composite materials. However, the most encouraging approach seems to be an integration of both living copolymerization and PIPS in one reactive media to deal with a soft material, not a glassy one. This approach is already in use when one speaks about dispersion polymerization and formation of copolymer aggregates.[4,5] Applying the same principles for bulk reactions and obtaining structured polymer matter directly in the reaction volume is an intriguing prospect of PIPS. There are several examples of such an approach in the recent literature, including synthesis of new ion-exchange[6,7] and porous[8] polymer materials, but usually the polymerization starts from complex presynthesized precursors and only one component is being polymerized.

Systems where both living copolymerization and structuring take place are very complex because the copolymer sequence space and the chains conformation space are being explored simultaneously.

---
[1] Corresponding author. E-mail: gavrilov@polly.phys.msu.ru

Thus the total phase space of such systems is very broad and difficult to study. To the best of our knowledge, one can mention only one paper by Kuchanov and Russo[9] where a theoretical approach was proposed to describe the copolymerization of monomer units with fixed reactivity ratios and additional Flory-Huggins χ-parameter to describe the interactions between monomers and growing copolymer chains. Unfortunately this approach has not been considered for polymer bulk, but only for dilute solutions and so-called globular nanoreactor formation.[10,11]

In addition we would like to mention several relevant and "emeritus" complex kinetic models of copolymerization, where the reactivity of the growing macroradical end depends on the macroradical sequence, namely penultimate[12] and even penpenultimate models,[13] see ref.[14] for a comprehensive review. However, these models are mainly focused on the sequence statistics and averaged properties rather than the description of the reaction kinetics coupled with the local surroundings. That is, all kinetic models do not consider possible heterogeneities in the reaction media.

In parallel to the development of new synthetic procedures and attempts of building analytical theory computer simulations have been widely used to simulate the polymerization processes. The studies and characterization of the properties of new materials have largely benefited from *in silico* experiments.[15,16,17] To the best of our knowledge, the very first attempt to simulate PIPS was described in the paper by Zhu[18] where Monte-Carlo simulations of PIPS during polymerization of one component was conducted in 2D. Later Lee[19,20] performed similar simulations using the classical Molecular Dynamics (MD) in a very thin layer. More recently the influence of polymerization on the phase separation of binary immiscible mixtures has been investigated by the Dissipative Particle Dynamics (DPD) simulations in two dimensions.[21] The authors observed complex phase separation behavior which was attributed to the interplay between the increasing thermodynamic driving force for phase separation and the increasing viscosity that suppresses phase separation during the polymerization process.

There are only few examples of works utilizing more rigorous and straightforward simulation approach explicitly taking into account the local surrounding in 3D. In the works of Genzer's group different types of surface and bulk initiated homopolymerization are studied; the MC simulation scheme based on the bond fluctuation model where the monomers and polymers reside on a three-dimensional cubic lattice was used to model a "living"/controlled radical polymerization.[22,23] Starovoitova et al. studied copolymerization of a single molecule near a selectively adsorbing surface and observed the formation of gradient copolymers.[24] Berezkin et al. simulated copolymerization of a single chain with a simultaneous globule formation,[25,26] which is in many senses similar to the analytical works of Kuchanov already mentioned above.[10,11] See refs. [27,28] for some recent comprehensive reviews in these fields. However, all these studies were mainly focused only on special and unusual conditions of copolymerization, such as a close vicinity to a flat surface or a single globular conformation of the growing chain. To the best of our knowledge there are no relevant works studying the bulk copolymerization process with simultaneous phase separation at a particle-based level. There are several well-known evidences of the so-called "bootstrap" effect,[29,30] when the growing polymer chain controls its own environment and the local species concentration significantly differs from the system-average levels.

Experimental results presented in Refs. [29,30] strongly disagreed with all the known kinetic models. Nevertheless, different kinetic models are still the most dominant way to perform computer experiments to study the copolymerization process. While these are not really *in silico* experiments, there are many tunable parameters and more or less any experimental result can be fitted by a kinetic model. Two approaches are commonly used to obtain polymer sequences using the kinetic models. The first is a mathematic-numeric approach based on a system of differential equations. Examples of its usage in the field of controlled radical copolymerization can be found in the works of Barner-Kowollik et al.[31] and Charleux et al.;[32] a popular commercial software PREDICI® by CiT (Computing in Technology, GmbH) was used in them. The second less often used approach is the application of Monte Carlo (MC) methods, in which polymer chains are simulated as individual objects. In contrast to the mathematic-numeric approach, MC simulations have a defined number of objects used in the model. If the chosen number is too small, the simulation will lead to incorrect results, while a large chain ensemble will increase the computation time and memory requirements. The ATRP copolymerization,[33] NMP copolymerization,[34] and gradient copolymerization with tracking the sequence distribution[35,36] are interesting examples of the application of the Monte Carlo simulation approach. We again would like to emphasize that both these approaches cannot take into account spatial heterogeneities and growing chain conformational properties. In other words, full compatibility of all the reactive species and products is assumed.

Thus, in this paper we are going to answer the following question: what is the influence of spatial heterogeneities caused by the species partial immiscibility on the growing chains sequences and overall system structure. Partly these questions have been already addressed in works [37,38], where we studied step-growth copolymerization in heterogeneous systems. Here we explore a more common and interesting case of the living radical copolymerization. We use Dissipative Particles Dynamics (DPD) with explicit initiator particles and local Monte-Carlo-like propagation step. Besides that the main assumptions will be the same miscibility of a free monomer particle and a corresponding monomer unit in a chain and the simplest terminal model (characterized by only two copolymerization rates, $r_1$ and $r_2$) for the chain propagation.

## 2. Simulation methodology

### 2.1. Dissipative particle dynamics method

First we give a brief description of the dissipative particle dynamics method. Dissipative particle dynamics (DPD) is a version of the coarse-grained molecular dynamics adapted to polymers and mapped onto the classical lattice Flory–Huggins theory.[39–42] It is a well-known method which has been utilized to simulate properties of a wide range of polymeric systems, such as single chains in solutions[43], polymer melts[44] and composites[45–47]. Macromolecules are represented in terms of the bead-and-spring model, with beads interacting by a conservative force (repulsion) $\boldsymbol{F}_{ij}^c$, a bond stretching force (only for connected beads) $F_{ij}^b$, a dissipative force (friction) $F_{ij}^d$, and a random force (heat generator) $F_{ij}^r$. The total force is given by:

$$F_i = \sum_{i \neq j} \left( F_{ij}^c + F_{ij}^b + F_{ij}^d + F_{ij}^r \right) \qquad (1)$$

The soft core repulsion between *i*- and *j*-th beads is equal to:

$$F_{ij}^c = \begin{cases} a_{\alpha\beta}(1 - r_{ij}/R_c)r_{ij}/r_{ij}, & r_{ij} \leq R_c \\ 0, & r_{ij} > R_c \end{cases}, \qquad (2)$$

where $r_{ij}$ is the vector between *i*-th and *j*-th bead, $a_{\alpha\beta}$ is the repulsion parameter if the particle *i* has the type *α* and the particle *j* has the type *β* and $R_c$ is the cutoff distance. $R_c$ is basically a free parameter depending on the volume of real atoms each bead represents;[42] $R_c$ is usually taken as the length scale, i.e. $R_c$=1. In our simulations we used $a_{\alpha\alpha}$=25. In this case, the interaction parameters $a_{\alpha\beta}$ and a more common Flory-Huggins parameter $\chi$ are linearly related to each other:[42]

$a_{\alpha\beta} = \chi/0.286 + 25$,  α≠β.

If two beads (*i* and *j*) are connected by a bond, there is also a simple spring force acting on them:

$$F_{ij}^b = -K(r_{ij} - l_0)\frac{r_{ij}}{r_{ij}}, \qquad (3)$$

where *K* is the bond stiffness and $l_0$ is the equilibrium bond length. We do not give here a more detailed description and parameters discussion of the standard DPD scheme, it can be found elsewhere.[42] In our simulations we used the following set of parameters: $a_{\alpha\alpha}$=25, *K*=4, $l_0$=0. The simulation box was set to be cubic with periodic boundary conditions in all three directions; we used 64x64x64 DPD units box (786432 beads) containing 786 initiators (thus resulting in the average chain length of ≈1000 at 100% conversion degree, which is a reasonable number from the experimental point of view), unless otherwise specified explicitly in the text.

## 2.2. Implementation of a polymerization reaction in DPD

The implementation of reactions in particle simulations is more or less common nowadays and several examples can be found in Refs. [23,37,48,49]. To simulate living polymerization we used a scheme similar to that described in ref. [23], with some simplifications described below. Every radical polymerization reaction characterized by 4 typical processes: initiation (i), chain growth - propagation (ii), chain transfer (iii) and termination (iv). While there are several different chemical schemes of living polymerization realization, there are always two very general features: fast initiation and negligible chain transfer.[1] In addition, since the equilibrium is always shifted to the dormant state, there are only a few active chain ends at any given moment and the chain termination via combination or disproportionation is very unlikely. Consequently, the whole ensemble of chains grows almost uniformly with molecular mass increasing proportionally to the conversion degree; another characteristic feature of a living polymerization process is a low polydispersity index (PDI) of the obtained polymer chains.

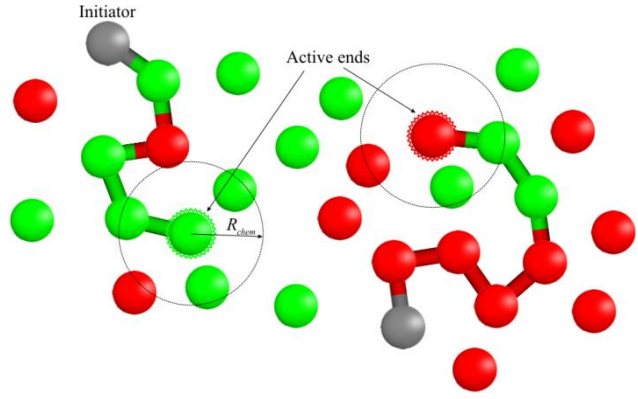

Fig. 1. Schematic illustration of the reaction scheme

To mimic the reaction process we use standard Monte Carlo scheme, the reaction procedure runs after each $\tau_0$ DPD steps. The reaction procedure consists of the following stages (see Fig. 1):

1) For each growing chain end, it becomes "dormant" with the probability $p_\alpha^e$ or "living" with the probability (1- $p_\alpha^e$), with $\alpha$ being the type of the chain end. Thus, the dormant/living active centers ratio is equal to $\tau_\alpha = p_\alpha^e/(1- p_\alpha^e)$.

2) Some growing chain end or initiator bead $i$ is selected at random;

3) If the selected active center is in the "dormant" state go to 2).

4) The list of all monomer beads located closer than the reaction radius $R_{chem}$ from the bead $i$ is created. The closest bead $j$ is determined, and a bond between the beads $i$ and $j$ is created with the probability $p_{ij}$. If the bond is not created, the procedure is repeated with the next closest bead from the list until a bond is created or there are no more unchecked monomers.

Stages 2) - 4) were repeated $M$ times where $M$ is the total number of the chain ends plus initiators, so that on average every growing chain is checked once every time we run the reaction procedure. The reaction radius $R_{chem}$ was chosen to be equal to 1.0, i.e. to the interaction potential cutoff distance $R_c$. For the reason of a very small probability of the chain termination and transfer processes during a living polymerization, we do not explicitly include them into our model.

We set the time interval between reaction steps $\tau_0 = 200$ DPD steps. This value is large enough to have local spatial equilibration in the nearest surrounding of each active center. At the same time, it is small enough to simulate a nearly continuous (non-discrete) process and to obtain high conversion degrees in a reasonable computational time.

Since we study a copolymerization process, there are two types of monomers, A and B (red and blue particles, see Figure 1) present in the system; the full reaction probability matrix is defined by 4 parameters: $p_{BB}$, $p_{AA}$, $p_{BA}$ and $p_{AB}$. Therefore, the expressions for the commonly used copolymerization constants are: $r_A = p_{AA}/p_{AB}$ and $r_B = p_{BB}/p_{BA}$. Since $r_A$ and $r_B$ are basically input parameters for our model, we still have to set 2 additional values, for example $p_{BA}$ and $p_{AB}$, to have the reaction probability matrix fully determined. In all our simulations we used $p_{BA}=p_{AB}=0.01$ because these probabilities define nothing but the overall reaction characteristic time. Since a living radical polymerization is a slow process, we want this characteristic time to be large compared to the diffusion characteristic time, i.e. we want the reaction to be kinetically controlled. Following the method for estimating these two characteristic times presented in ref.[50] we found that the value of $p_{BA}=p_{AB}=0.01$ in conjunction with $\tau_0 = 200$ DPD steps is small enough to simulate a kinetically controlled reaction.

Another two parameters of the copolymerization scheme are the dormant/living active centers ratio for two types of chain ends (A and B): $\tau_A$ and $\tau_B$. The monomer feed is characterized by the A-monomer volume fraction $\varphi_A=1-\varphi_B$.

## 2.3 Monte Carlo calculations

In addition to DPD simulations, we used a simple Monte Carlo model in the sequence space. The first monomer in the chain is chosen according to the monomer volume fractions $\varphi_A$ and $\varphi_B$; the propagation probabilities depend on $r_A$ and $r_B$ as follows: $p_{AA}^{MC} = \dfrac{r_A \varphi_A}{1-\varphi_A + r_A \varphi_A}$, $p_{AB}^{MC} = \dfrac{1-\varphi_A}{1-\varphi_A + r_A \varphi_A}$, $p_{BB}^{MC} = \dfrac{r_B \varphi_B}{1-\varphi_B + r_B \varphi_B}$, $p_{BA}^{MC} = \dfrac{1-\varphi_B}{1-\varphi_B + r_B \varphi_B}$, thus representing a simple markovian process. In this model we additionally took into account that the reaction volume is finite and the monomer volume fractions change during the copolymerization process; the initial number of monomers and initiators in every Monte Carlo calculation was equal to that in the corresponding DPD system. This model is used as an additional tool to verify the correctness of the results of DPD simulations.

## 2.4. Parameters choice

As it was mentioned in the previous section, every polymerization process in our model has 4 principal parameters: two copolymerization degrees $r_A$ and $r_B$ and two dormant/living active centers ratios $\tau_A$ and $\tau_B$. While the former two parameters have an obvious impact on the chain sequence of the resulting polymer, it is unclear how the latter two influence the chain growth. In order to test that, we simulated a trial copolymerization process with fixed $r_A = r_B = 0.5$ (i.e. symmetrical process with both $r_A$ and $r_B < 1$ which is typical for radical copolymerization[1]) but with three different sets of $\tau_A$ and $\tau_B$: 1) $\tau_A = \tau_B = 0$, i.e. no dormant ends in the system; 2) $\tau_A = 9$ and $\tau_B = 0$, i.e. no dormant ends of type B but on average 90% ends of type A are in the dormant state; 3) $\tau_A = 0$ and $\tau_B = 9$, i.e. no dormant ends of type A but on average 90% ends of type B are in the dormant state. The A-monomer volume fraction was equal to $\varphi_A = 0.3$ and the Flory-Huggins parameter was equal to $\chi = 1.8$. The choice of $\varphi_A$ was dictated by the fact that we wanted to study the influence of dormant ends for both major and minor fractions. We did not study the case of $\tau_A = 9$ and $\tau_B = 9$ because the copolymer sequence would obviously be the same as for the case of $\tau_A = \tau_B = 0$.

To study the resulting polymers, we calculated the dependences of PDI and the triad fractions on the conversion degree. The former value gives us some information about the ensemble of chains in general, while the latter values are often used to characterize the chain sequences.[29] Results are presented in Fig. 2.

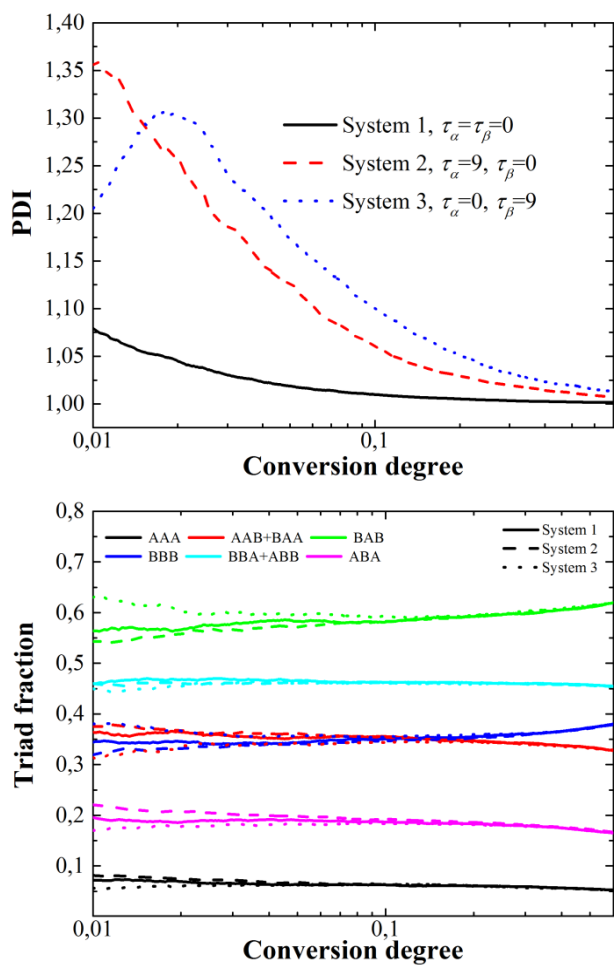

Fig. 2 The dependence of PDI (top) and triad fractions (bottom) for three systems with different amount of dormant ends. The A-centered and B-centered triad fraction sets are normalized so that the sums are both equal to 1.0.

One can see that the dependences of PDI on conversion degree look rather different, especially at low conversion degrees. It is to be expected, because due to the presence of dormant ends not all the chains are growing simultaneously but only the fraction being in the living state, which obviously leads to an increase in PDI. However, this effect becomes much less pronounced with growing conversion degree because on average all the chains stay in the living state exactly the same time.

Examination of the triad fractions reveals that the chain sequences are indistinguishable starting from rather small conversion degrees of approximately 5-6%, i.e. from the stage where the average chain length is much larger than 3 (i.e. the triad length) and the influence of larger PDI for systems 2 and 3 becomes negligible. Therefore, we can conclude that explicit accounting for the dormant ends does not change the chain sequences, and further in our simulations we will use $\tau_A = \tau_B = 0$ and study the systems at 10% conversion degree to be sure that the polymer chains are long enough for PDI not to affect the sequences. The obtained results also confirm that our choice to fix $p_{BA} = p_{AB}$ does not influence the copolymer sequences, because changing $p_{AB}$ with fixed $r_A$ (or $p_{BA}$ with fixed $r_B$) is on average equivalent to a corresponding change in $\tau_A$ (or $\tau_B$).

## 3. Results

### 3.1 Styrene-acrylic acid copolymerization: comparison with the available literature data, Mayo-Lewis theory and Monte-Carlo calculations

To verify our model in more detail in this section we examine the bulk radical copolymerization of a well-studied system of styrene (S) - acrylic acid (AA) copolymer (PS-PAA). We would like to note that several slightly different polymerization constants for the PS-PAA system could be found in the literature, and the particular values depend on the solvent and mechanism of the living process. For instance, in the work [51] the NMP process was realized and the following reactivity ratios were determined: $r_A = 0.27$ and $r_S = 0.72$; in the work [52] the RAFT polymerization was used and the authors obtained $r_A=0.082$ and $r_S=0.21$.

In this subsection we used the copolymerization constants reported in the work [53]: $r_A = p_{AA}/p_{AS} = 0.13$ and $r_S = p_{SS}/p_{SA} = 0.38$. While the non-living process was utilized in that work, the resulting copolymers were extensively studied using $^1$H and $^{13}$C NMR experiments to obtain the triad fractions, which are an excellent target for comparison with the results of simulations. One might speculate that if the conversion degree and the initiator concentration are low enough the chain sequences obtained using living and non-living copolymerization would be the same; keeping this in mind, we compared the results of our living polymerization model with the experimental results obtained in the work [53]. We assumed that the incompatibility parameter $\chi=0$ because the experimental copolymerization constants are obtained from the models that disregard the presence of incompatibility between AA and S.

The most common way to characterize a copolymerization process is the so-called "composition-feed curve": the average composition of the obtained polymer is presented as a function of the initial monomer feed composition; see Fig. 3. One can see that the PS-PAA composition-feed curve is a classical example of an S-shaped curve for the case when the both reaction rates $r_A$, $r_S < 1.0$. Both the simulated data and the Monte Carlo results agree nicely with the experimental data from ref.[53].

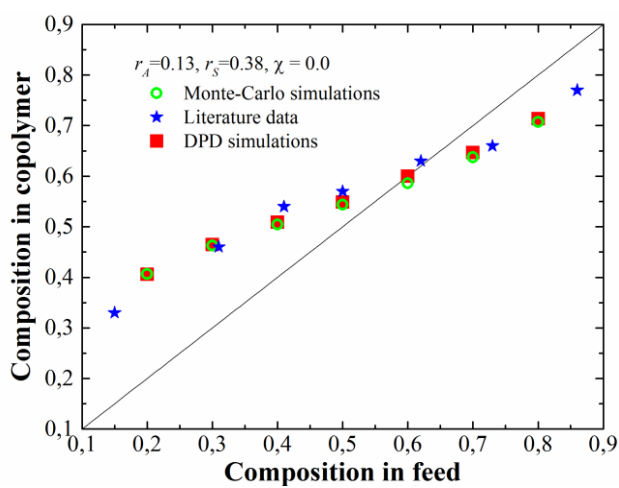

Fig. 3 The composition-feed curves obtained from DPD simulations, Monte-Carlo calculations and experimental data (ref. [53]).

The triad composition was studied next. The triad fractions values obtained from DPD simulations, Mayo-Lewis theory, Monte Carlo calculations in comparison with the experimental data obtained by $^1$H and $^{13}$C NMR experiments[53] for $\varphi_A$=0.5 are presented in Table 1. We can see some minor deviations of the experimental data from the theory and calculations, which can be explained by either experimental uncertainties or the influence of monomers incompatibility; we study the influence of incompatibility on the chain sequences in the next subsection.

Table 1. Comparison of triad fractions for PS-PAA copolymerization obtained by different methods at $\varphi_A$=0.5.

| PS-PAA | Data type | $f_{AAA}$ | $f_{AAS}$ | $f_{SAS}$ | $f_{SSS}$ | $f_{SSA}$ | $f_{ASA}$ |
|---|---|---|---|---|---|---|---|
| $r_1$ = 0.13 $r_2$ = 0.38 | Mayo-Lewis eqations[53] | 0.01 | 0.20 | 0.79 | 0.08 | 0.40 | 0.51 |
| | Monte-Carlo calculation | 0,013 | 0,204 | 0,783 | 0,074 | 0,396 | 0,530 |
| | NMR Experiments[53], $\varphi_A$=0.491 | 0.024 | 0.207 | 0.768 | 0.109 | 0.435 | 0.456 |
| | DPD simulation | 0.016 | 0.203 | 0.781 | 0.074 | 0.398 | 0.528 |

**3.2 Influence of spatial inhomogeneities**

From the previous section we can conclude that the DPD model yields correct composition-feed curves and sequences and can be used further to investigate the influence of a nonzero χ on the obtained polymer structure. In this subsection we will again consider an abstract case of a model symmetric reaction between A and B monomers with $r_A = r_B = 0.5$. First of all we plot the copolymer composition curve at two very different incompatibilities: at χ = 0.0 and χ = 1.8, see Fig. **4**. These two values represent two limiting cases of fully compatible system (χ = 0.0) and very incompatible system (χ = 1.8), which is close to the critical point.

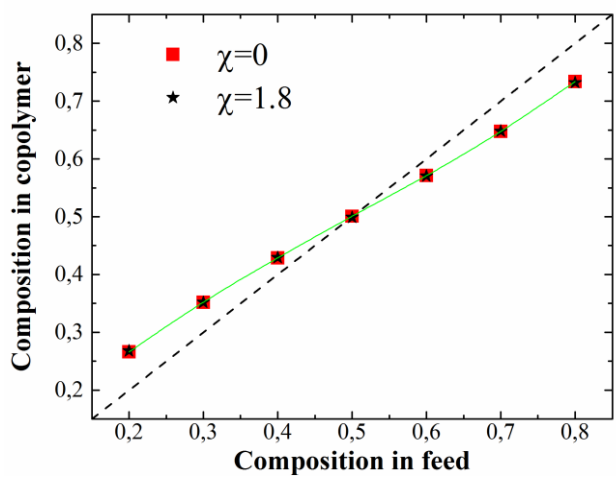

Fig. 4. The composition-feed curves for model reaction with $r_A = r_B = 0.5$ at different χ-values.

Surprisingly we do not see any deviations in composition at all, the calculated composition value at $\chi = 1.8$ coincide with the corresponding values at $\chi = 0.0$. This result looks counter-intuitive, initially we expected at least some differences between these two limiting cases. In addition we do not see a remarkable difference in the 3D snapshots at different $\chi$-values (not presented here), both before and after the polymerization process.

While there are no changes in the composition profile, we can expect to find some differences at the more sophisticated level of sequence statistics. First of all, we can calculate the "observable" reaction rates $r'$ from a direct analysis of the growing sequences. That is, we can calculate the real probabilities to find two consequential A-units, two consequential B-units, A-unit after B-unit and vise-versa (i.e. $p'_{AA}$, $p'_{BB}$, $p'_{AB}$ and $p'_{BA}$) by "scanning" along the chains and then obtain $r'_A = p'_{AA}/p'_{AB}$ and $r'_B = p'_{BB}/p'_{BA}$. We distinguish these "observable" $r'_A$ and $r'_B$ from the fixed input reaction rate parameters $r_A$ and $r_B$ because they take into account the local surrounding of the growing chain and possible heterogeneities.

There is one more way to obtain the reaction rates from the chain sequence – calculate them directly from the triad fractions using the Mayo-Lewis equations. The reaction rates $r^*_A$ and $r^*_B$ at small conversions could be easily obtained from the known dependencies for symmetric triads:[53,54] $f_{AAA} = P_{AA}^2$, $f_{BBB} = P_{BB}^2$, $f_{ABA} = (1 - P_{BB})^2$, $f_{BAB} = (1 - P_{AA})^2$, where $P_{AA} = r_A \varphi_A / (1 - \varphi_A + r_A \varphi_A)$, $P_{BB} = r_B \varphi_B / (1 - \varphi_B + r_B \varphi_B)$. From these equations we have: $r^*_A = \dfrac{(1-\varphi_A)\sqrt{f_{AAA}}}{\varphi_A(1-\sqrt{f_{AAA}})}$ from AAA triads, $r^*_B = \dfrac{(1-\varphi_B)\sqrt{f_{BBB}}}{\varphi_B(1-\sqrt{f_{BBB}})}$ from BBB triads, $r^*_A = \dfrac{(1-\varphi_A)(1-\sqrt{f_{BAB}})}{\varphi_A \sqrt{f_{BAB}}}$ from BAB triads and $r^*_B = \dfrac{(1-\varphi_B)(1-\sqrt{f_{ABA}})}{\varphi_B \sqrt{f_{ABA}}}$ from ABA triads. These "triad" rates will help us to illustrate what reaction rates values are obtained if one applies the standard Mayo-Lewis equations of homogeneous copolymerization to a process with spatial inhomogeneities.

Thus, in addition to the fixed input parameters of our simulation ($r_1$ and $r_2$) we studied two more sets of values: the "observable" $r'_A$ and $r'_B$ and the "triad" rates $r^*_A$ and $r^*_B$. The plots of these reaction rates versus $\chi$ parameter and feed composition are presented in Fig. 5. One can see clear differences between the preset, "observable" and "triad" reaction rates calculated from different types of triads (again, during all the simulations the preset reaction rates were fixed at $r_A = r_B = 0.5$). First of all, both the "observable" and "triad" reaction rates grow upon increasing $\chi$ (see Fig. 5, top); the increase itself is about 50% at $\chi = 1.8$. Moreover, the rates $r^*$ calculated from homo-triads (AAA and BBB) overestimate the "observable" $r'$ values, while the values obtained from hetero-triads ABA and BAB underestimate them; this deviation becomes more pronounced at larger $\chi$ values. Second, the "observable" reaction rates $r'$ depend strongly on the feed composition (see Fig. 5, bottom): the reaction rates $r'_A$ and $r'_B$ increase with the increase in the corresponding values of monomer volume fractions, $\varphi_A$ and $\varphi_B$.

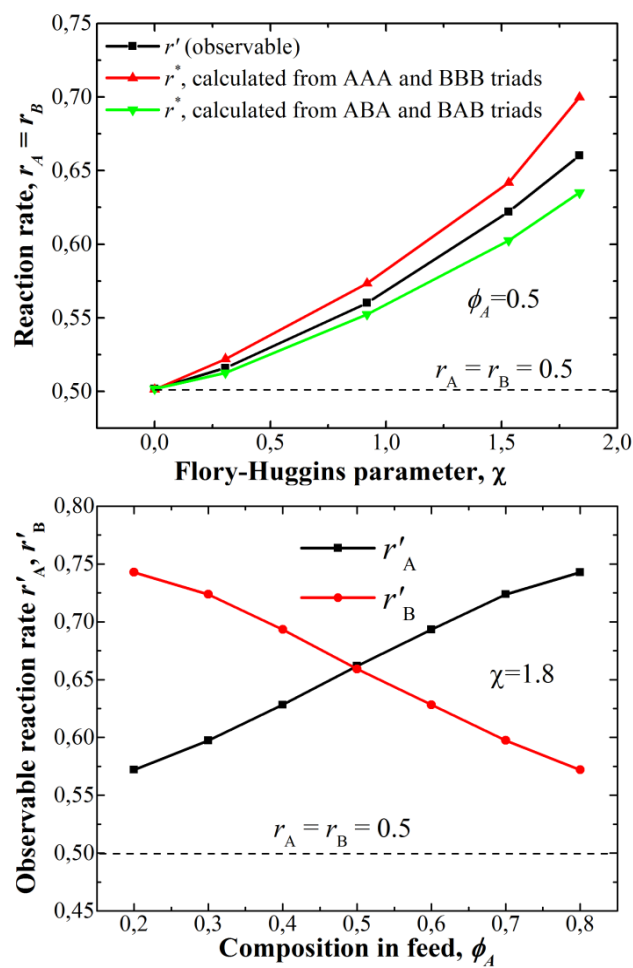

Fig. 5 Reaction rates ("observable" $r'$ and "triad" $r^*$) versus $\chi$ parameter (top) and feed composition $\varphi$ (bottom).

In principle all the results presented in Fig. 5 look reasonable and can be explained by the presence of spatial monomer concentration inhomogeneities caused by a large $\chi$-value. However, it is still unclear why the same mechanism does not affect the composition-feed curve. We suppose that a possible explanation for this (independence of the composition but strong differences in the sequence statistics) is the fact that the reaction is indeed kinetically controlled. At high $\chi$-values the concentration fluctuations are increased, which results in the aforementioned *local* changes in the chain sequences. On the contrary, the chain composition is an integral value, and on average each chain grows in a homogeneous environment because the local concentration changes are significantly faster than the chain growth process. Thus, the average copolymer composition stays the same, but the observable reactivity ratios change significantly with a remarkable increase of the chain blockiness. Therefore, we observed the bootstrap effect in our system.

In our study both reactivity rates $r_A$ and $r_B$ were below 1.0 (the most common case for radical copolymerization) which resulted in obtaining alternating sequences even for very high $\chi$-values: the "observable" reactions rates were still smaller than 1.0 even at the highest studied $\chi$-values; this is the reason why no microphase separation was observed. This observation is valid not only for small conversion degrees, but also for moderate and high conversion degrees in the case of a symmetrical feed composition. However, highly unsymmetrical feed compositions could lead to more complex sequences at high conversions, in particular gradient copolymers, which will be studied next.

### 3.3. Gradient copolymers

In this section we test our model on a system where one can expect a microphase separation during copolymerization. A representative case of gradient copolymers synthesis was described in ref. [52] where gradient copolymers were obtained for the same styrene - acrylic acid pair studied in section 3.1. It was shown that during a RAFT-mediated living process starting from an unsymmetrical feed with styrene minor fraction gradient block-like copolymers are formed; these copolymers consist of a random copolymer segment which grows first, a short transitional segment and an almost pure acrylic-acid segment.[52] This happens because $r_A < r_S$ which in conjunction with $\varphi_S < \varphi_A$ leads to a sharp exhaustion of the styrene monomer at some intermediate conversion degree. We simulated that process for $\varphi_S = 0.2$ using the copolymerization constants reported in ref. [52]: $r_S = 0.21$ and $r_A = 0.081$. We used the same 64x64x64 DPD units box but this time containing 7864 initiators to reduce the average chain length at 100% conversion degree to 100 monomer units. Two limiting cases were studied: the ideal case of $\chi = 0$ and the case of a very large incompatibility between S and AA monomers, $\chi = 1.9$, which is close to the critical point of monomer-monomer segregation $\chi = 2.0$. The latter choice is dictated by the fact that while there is no clear information on the real value of $\chi$-parameter between S and AA monomer units in the bulk, we wanted to investigate whether microphase separation with long-range order during polymerization is possible in this system in general. First, we studied the distributions of monomer units along the chain at 99.9% conversion degree; the result is shown in Fig. 6.

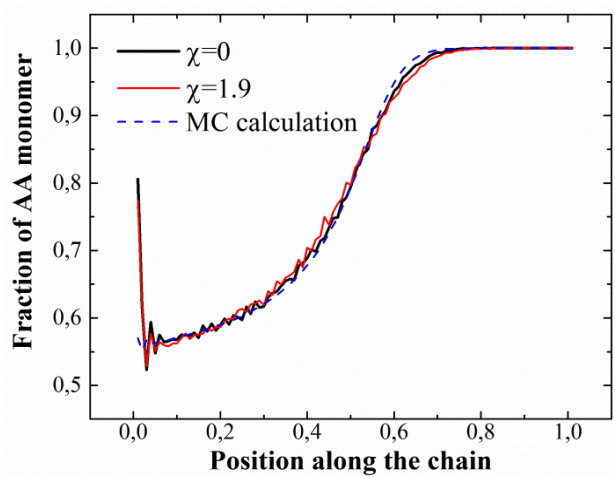

Fig. 6 Fraction profiles of AA monomer along the chain obtained from DPD simulations at $\chi = 0.0$ and $\chi = 1.9$ in comparison with the results of Monte Carlo calculation. The DPD results are obtained by averaging over all the chains in the system and normalizing their length to 1.0.

As one can see, gradient copolymers are indeed synthesized during the reaction. No differences between the systems with $\chi = 0$ and 1.9 are observed, but it was to be expected as the curves in Fig. **6** reflect the compositional changes along the chain, and we previously showed that increasing $\chi$ has no effect on the chain composition. The differences in the chain sequences statistics are the same as described in the previous section and we therefore do not discuss them here. It is worth noting that a perfect agreement with the Monte Carlo calculation is observed neglecting the fact that the profiles obtained from DPD simulations are slightly smoother due to the averaging over chains with different lengths. Fig. 7 shows a visual representation of all the chains present in the system with $\chi = 1.9$, which again confirms that block-like polymers are formed.

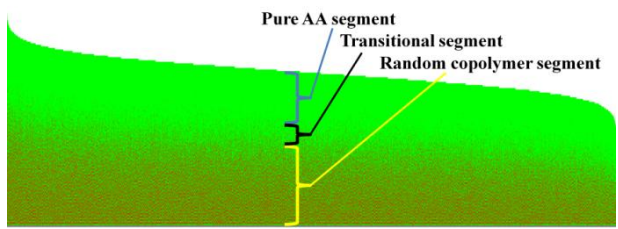

Fig. 7 Visual representation of the chain ensemble for the system with $\chi = 1.9$. The initiators are depicted in grey, the S monomer units are depicted in red and the AA monomer units are green. The chains are sorted according to their length for the visual clarity.

Finally, we studied the microstructure of the system, see Fig. 8. It is obvious that at $\chi = 0.0$ no segregation was observed (see Fig. 8, top), while for $\chi = 1.9$ a well-defined lamellar structure was obtained (see Fig. 8, bottom). Therefore, in our model we indeed can observe a phase separation during undergoing copolymerization.

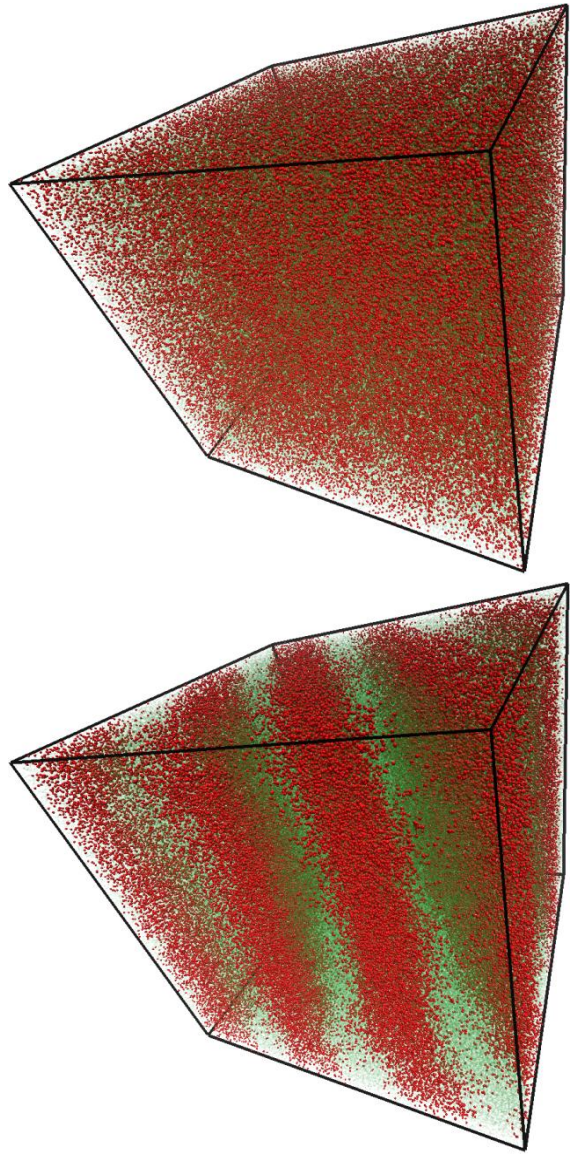

Fig. 8 Lamellar structure observed for $\chi = 0.0$ (top) and $\chi = 1.9$ (bottom), S-units shown in red, while AA units are shown in green and are semi-transparent.

While the incompatibility is very high, $\chi N = 190$, the degree of segregation of the resulting structure is not, meaning there are no sharp phase boundaries. This can be explained by the fact that the first segment of the resulting "pseudo-diblock" copolymers contains large amount of AA monomers, which effectively decreases the segregation between the first and the third segments (see Fig. 7). It is interesting that there is no changes in the monomer distribution along the chain (Fig. 6) even though a PIPS occurs for the system with $\chi = 1.9$. However, it is easy to understand because in order for the segregation to happen a pure AA segment of some length must be formed, which means that the residual amount of S monomer is very small and the chains will grow by addition of AA monomers independently of whether there is a phase separation or not.

## 4. Conclusions

In this paper we introduced a computer simulation methodology to study living radical heterogeneous copolymerization. In our model the spatial segregation of species is accounted for and a microphase separation could occur during the copolymerization process. We performed model verification on the well-studied system of styrene - acrylic acid bulk copolymerization, both for low and high conversion degrees. Our simulations exhibited a very good agreement with the available literature data and, moreover, allowed us to predict reactive melt 3D structure during the copolymerization process, including long-range ordered states during PIPS.

In addition we simulated a model symmetric reaction with various incompatibility parameters $\chi$ and studied the emerged difference between real, observable and calculated (from the assumption that the system is homogeneous) reaction rate values. We showed that at a low conversion degree the spatial inhomogeneties lead to an increased blockiness of the growing sequences, while there is no any influence on the copolymer composition even at very high $\chi$-values. This effect could be explained that on average each chain grows in a homogeneous environment because the local concentration changes significantly faster compared to the chain growth process. Therefore the average copolymer composition stays the same, but the observable reactivity ratios change significantly.

The described methodology can be applied to study polymerization in initially heterogeneous systems (like emulsion polymerization) without any significant modifications. Another interesting process to be investigated is ionic copolymerization, in which rather big values of copolymerization constants could be observed. However, some modifications to the method should be made to take into account complex processes occurring during that kind of copolymerization.

**Acknowledgements**

This work was done within the Russian Science Foundation project 14-13-00683. The authors thank M.Zaremskii for critical reading and comments about living copolymerization schemes and Moscow State University Supercomputer Center[55] for providing the computational resources.

**References**

(1) Braunecker, W. A.; Matyjaszewski, K. Controlled/living Radical Polymerization: Features, Developments, and Perspectives. *Progress in Polymer Science (Oxford)*. 2007, pp 93–146.

(2) Inoue, T. Reaction-Induced Phase Decomposition in Polymer Blends. *Prog. Polym. Sci.* **1995**, *20* (1), 119–153 DOI: 10.1016/0079-6700(94)00032-W.

(3) Liu, Y. Polymerization-Induced Phase Separation and Resulting Thermomechanical Properties of Thermosetting/reactive Nonlinear Polymer Blends: A Review. *Journal of Applied Polymer Science*. 2013, pp 3279–3292.

(4) Derry, M. J.; Fielding, L. A.; Armes, S. P. Polymerization-Induced Self-Assembly of Block Copolymer Nanoparticles via RAFT Non-Aqueous Dispersion Polymerization. *Progress in Polymer Science*. 2016, pp 1–18.

(5) Sun, J.-T.; Hong, C.-Y.; Pan, C.-Y. Formation of the Block Copolymer Aggregates via

Polymerization-Induced Self-Assembly and Reorganization. *Soft Matter* **2012**, *8* (30), 7753 DOI: 10.1039/c2sm25537e.

(6)  McIntosh, L. D.; Schulze, M. W.; Irwin, M. T.; Hillmyer, M. A.; Lodge, T. P. Evolution of Morphology, Modulus, and Conductivity in Polymer Electrolytes Prepared via Polymerization-Induced Phase Separation. *Macromolecules* **2015**, *48* (5), 1418–1428 DOI: 10.1021/ma502281k.

(7)  Chopade, S. A.; So, S.; Hillmyer, M. A.; Lodge, T. P. Anhydrous Proton Conducting Polymer Electrolyte Membranes via Polymerization-Induced Microphase Separation. *ACS Appl. Mater. Interfaces* **2016**, *8* (9), 6200–6210 DOI: 10.1021/acsami.5b12366.

(8)  Oh, J.; Seo, M. Photoinitiated Polymerization-Induced Microphase Separation for the Preparation of Nanoporous Polymer Films. *ACS Macro Lett.* **2015**, *4* (11), 1244–1248 DOI: 10.1021/acsmacrolett.5b00734.

(9)  Kuchanov, S. I.; Russo, S. Quantitative Theory of Free-Radical Copolymerization Allowing for the Phenomenon of Preferential Sorption. *Macromolecules* **1997**, *30* (16), 4511–4519 DOI: 10.1021/ma961473m.

(10) Kuchanov, S. I.; Pogodin, S. G.; Ten Brinke, G.; Khokhlov, A. Polymer Globule as a Nanoreactor. *Macromolecules* **2008**, *41* (7), 2689–2693 DOI: 10.1021/ma702710s.

(11) Kuchanov, S. I.; Pogodin, S. G. Theoretical Consideration of Bulk Free-Radical Copolymerization with Allowance for the Preferential Sorption of Monomers into Globular Nanoreactors. *J. Chem. Phys.* **2008**, *128* (24) DOI: 10.1063/1.2940351.

(12) Merz, E.; Alfrey, T.; Goldfinger, G. Intramolecular Reactions in Vinyl Polymers as a Means of Investigation of the Propagation Step. *J. Polym. Sci.* **1946**, *1* (2), 75–82 DOI: 10.1002/pol.1946.120010202.

(13) Ham, G. E. Expansion of Copolymerization Theory (Influence of Neighboring Units on Radical Reactivity with Monomers). *J. Polym. Sci.* **1960**, *45* (145), 169–175 DOI: 10.1002/pol.1960.1204514516.

(14) Coote, M. L.; Davis, T. P. Mechanism of the Propagation Step in Free-Radical Copolymerisation. *Prog. Polym. Sci.* **1999**, *24* (9), 1217–1251 DOI: 10.1016/S0079-6700(99)00030-1.

(15) Liu, A. J.; Grest, G. S.; Marchetti, M. C.; Grason, G. M.; Robbins, M. O.; Fredrickson, G. H.; Rubinstein, M.; Olvera de la Cruz, M. Opportunities in Theoretical and Computational Polymeric Materials and Soft Matter. *Soft Matter* **2015**, *11* (12), 2326–2332 DOI: 10.1039/C4SM02344G.

(16) Potestio, R.; Peter, C.; Kremer, K. Computer Simulations of Soft Matter: Linking the Scales. *Entropy*. 2014, pp 4199–4245.

(17) Peter, C.; Kremer, K. Soft Matter , Fundamentals and Coarse Graining Strategies. *Multiscale Simul. Methods Mol. Sci.* **2009**, *42*, 337–358.

(18) Zhu, Y.-M. Monte Carlo Simulation of Polymerization-Induced Phase Separation. *Phys. Rev. E* **1996**, *54* (2), 1645–1651 DOI: 10.1103/PhysRevE.54.1645.

(19) Lee, J. C. Polymerization-Induced Phase Separation. **1999**, *60* (2), 1930–1935.

(20) Lee, J. C. Polymerization-Induced Phase Separation: Intermediate Dynamics. *Int. J. Mod. Phys. C*

**2000**, *11* (2), 347–358 DOI: 10.1142/S0129183100000328.

(21)  Liu, H.; Qian, H.-J.; Zhao, Y.; Lu, Z.-Y. Dissipative Particle Dynamics Simulation Study on the Binary Mixture Phase Separation Coupled with Polymerization. *J. Chem. Phys.* **2007**, *127* (2007), 144903 DOI: 10.1063/1.2790005.

(22)  Turgman-Cohen, S.; Genzer, J. Simultaneous Bulk- and Surface-Initiated Controlled Radical Polymerization from Planar Substrates. *J. Am. Chem. Soc.* **2011**, *133* (44), 17567–17569 DOI: 10.1021/ja2081636.

(23)  Genzer, J. In Silico Polymerization: Computer Simulation of Controlled Radical Polymerization in Bulk and on Flat Surfaces. *Macromolecules* **2006**, *39* (20), 7157–7169 DOI: 10.1021/ma061155f.

(24)  Starovoitova, N. Y.; Berezkin, A. V.; Kriksin, Y. A.; Gallyamova, O. V.; Khalatur, P. G.; Khokhlov, A. R. Modeling of Radical Copolymerization near a Selectively Adsorbing Surface: Design of Gradient Copolymers with Long-Range Correlations. *Macromolecules* **2005**, *38* (6), 2419–2430 DOI: 10.1021/ma0487094.

(25)  Berezkin, A. V.; Khalatur, P. G.; Khokhlov, A. R. Computer Modeling of Synthesis of Proteinlike Copolymer via Copolymerization with Simultaneous Globule Formation. *J. Chem. Phys.* **2003**, *118* (17), 8049–8060 DOI: 10.1063/1.1563603.

(26)  Berezkin, A. V.; Khalatur, P. G.; Khokhlov, A. R.; Reineker, P. Molecular Dynamics Simulation of the Synthesis of Protein-like Copolymers via Conformation-Dependent Design. *New J. Phys.* **2004**, *6*, 44 DOI: 10.1088/1367-2630/6/1/044.

(27)  Genzer, J.; Khalatur, P. G.; Khokhlov, A. R. Conformation-Dependent Design of Synthetic Functional Copolymers. In *Polymer Science: A Comprehensive Reference, 10 Volume Set*; 2012; Vol. 6, pp 689–723.

(28)  Bain, E. D.; Turgman-Cohen, S.; Genzer, J. Progress in Computer Simulation of Bulk, Confined, and Surface-Initiated Polymerizations. *Macromolecular Theory and Simulations*. 2013, pp 8–30.

(29)  Harwood, H. J. Structures and Compositions of Copolymers. *Makromol. Chemie. Macromol. Symp.* **1987**, *10–11* (1), 331–354 DOI: 10.1002/masy.19870100117.

(30)  Semchikov, Y. D.; Slavnitskaya, N. N.; Smirnova, L. A.; Sherstyanykh, V. I.; Sveshnikova, T. G.; Borina, T. I. The Influence of Preferential Sorption upon the Copolymerization of Vinylpyrrolidone with Vinyl Acetate. *Eur. Polym. J.* **1990**, *26* (8), 889–891 DOI: 10.1016/0014-3057(90)90163-X.

(31)  Barner-Kowollik, C.; Quinn, J. F.; Morsley, D. R.; Davis, T. P. Modeling the Reversible Addition-Fragmentation Chain Transfer Process in Cumyl Dithiobenzoate-Mediated Styrene Homopolymerizations: Assessing Rate Coefficients for the Addition-Fragmentation Equilibrium. *J. Polym. Sci. Part A Polym. Chem.* **2001**, *39* (9), 1353–1365 DOI: 10.1002/pola.1112.

(32)  Nicolas, J.; Mueller, L.; Dire, C.; Matyjaszewski, K.; Charleux, B. Comprehensive Modeling Study of Nitroxide-Mediated Controlled/living Radical Copolymerization of Methyl Methacrylate with a Small Amount of Styrene. *Macromolecules* **2009**, *42* (13), 4470–4478 DOI: 10.1021/ma900515v.

(33) Al-Harthi, M. A.; Masihullah, J. K.; Abbasi, S. H.; Soares, J. B. P. Dynamic Monte Carlo Simulation of ATRP in a Batch Reactor. *Macromol. Theory Simulations* **2009**, *18* (6), 307–316 DOI: 10.1002/mats.200900001.

(34) Wang, L.; Broadbelt, L. J. Kinetics of Segment Formation in Nitroxide-Mediated Controlled Radical Polymerization: Comparison with Classic Theory. *Macromolecules* **2010**, *43* (5), 2228–2235 DOI: 10.1021/ma9019703.

(35) Wang, L.; Broadbelt, L. J. Factors Affecting the Formation of the Monomer Sequence along Styrene/methyl Methacrylate Gradient Copolymer Chains. *Macromolecules* **2009**, *42* (21), 8118–8128 DOI: 10.1021/ma901552a.

(36) Wang, L.; Broadbelt, L. J. Explicit Sequence of Styrene/methyl Methacrylate Gradient Copolymers Synthesized by Forced Gradient Copolymerization with Nitroxide-Mediated Controlled Radical Polymerization. *Macromolecules* **2009**, *42* (20), 7961–7968 DOI: 10.1021/ma901298h.

(37) Gavrilov, A. A.; Guseva, D. V.; Kudryavtsev, Y. V.; Khalatur, P. G.; Chertovich, A. V. Simulation of Phase Separation in Melts of Reacting Multiblock Copolymers. *Polym. Sci. Ser. A* **2011**, *53* (12), 1207–1216 DOI: 10.1134/S0965545X11120054.

(38) Gavrilov, A. A.; Chertovich, A. V. Self-Assembly in Thin Films during Copolymerization on Patterned Surfaces. *Macromolecules* **2013**, *46* (11), 4684–4690 DOI: 10.1021/ma4003243.

(39) Hoogerbrugge, P. J.; Koelman, J. M. V. A. Simulating Microscopic Hydrodynamic Phenomena with Dissipative Particle Dynamics. *Europhys. Lett.* **1992**, *19* (3), 155–160 DOI: 10.1209/0295-5075/19/3/001.

(40) Schlijper, A. G.; Hoogerbrugge, P. J.; Manke, C. W. Computer Simulation of Dilute Polymer Solutions with the Dissipative Particle Dynamics Method. *J. Rheol.* **1995**, *39* (3), 567–579 DOI: 10.1122/1.550713.

(41) Español, P.; Warren, P. Statistical Mechanics of Dissipative Particle Dynamics. *Europhys. Lett.* **1995**, *30* (4), 191–196 DOI: 10.1209/0295-5075/30/4/001.

(42) Groot, R. D.; Warren, P. B. Dissipative Particle Dynamics: Bridging the Gap between Atomistic and Mesoscopic Simulation. *J. Chem. Phys.* **1997**, *107* (11), 4423–4435 DOI: 10.1063/1.474784.

(43) Guo, J.; Liang, H.; Wang, Z.-G. Coil-to-Globule Transition by Dissipative Particle Dynamics Simulation. *J. Chem. Phys.* **2011**, *134* (24), 244904 DOI: 10.1063/1.3604812.

(44) Gavrilov, A. A.; Kudryavtsev, Y. V.; Chertovich, A. V. Phase Diagrams of Block Copolymer Melts by Dissipative Particle Dynamics Simulations. *J. Chem. Phys.* **2013**, *139* (22), 224901 DOI: 10.1063/1.4837215.

(45) Raos, G.; Casalegno, M. Nonequilibrium Simulations of Filled Polymer Networks: Searching for the Origins of Reinforcement and Nonlinearity. *J. Chem. Phys.* **2011**, *134* (5), 54902 DOI: 10.1063/1.3537971.

(46) Gavrilov, A. A.; Chertovich, A. V.; Khalatur, P. G.; Khokhlov, A. R. Study of the Mechanisms of Filler Reinforcement in Elastomer Nanocomposites. *Macromolecules* **2014**, *47* (15), 5400–5408

DOI: 10.1021/ma500947g.

(47) Gavrilov, A. A.; Komarov, P. V.; Khalatur, P. G. Thermal Properties and Topology of Epoxy Networks: A Multiscale Simulation Methodology. *Macromolecules* **2015**, *48* (1), 206–212 DOI: 10.1021/ma502220k.

(48) Glagolev, M. K.; Lazutin, A. A.; Vasilevskaya, V. V.; Khokhlov, A. R. Influence of Cross-Linking Rate on the Structure of Hypercrosslinked Networks: Multiscale Computer Simulation. *Polymer* **2016**, *86*, 168–175 DOI: 10.1016/j.polymer.2016.01.040.

(49) Yong, X.; Kuksenok, O.; Balazs, A. C. Modeling Free Radical Polymerization Using Dissipative Particle Dynamics. *Polymer* **2015**, *72*, 217–225 DOI: 10.1016/j.polymer.2015.01.052.

(50) Berezkin, A. V.; Kudryavtsev, Y. V. Simulation of End-Coupling Reactions at a Polymer-Polymer Interface: The Mechanism of Interfacial Roughness Development. *Macromolecules* **2011**, *44* (1), 112–121 DOI: 10.1021/ma101285m.

(51) Couvreur, L.; Charleux, B.; Guerret, O.; Magnet, S. Direct Synthesis of Controlled Poly(styrene-Co-Acrylic Acid)s of Various Compositions by Nitroxide-Mediated Random Copolymerization. *Macromol. Chem. Phys.* **2003**, *204* (17), 2055–2063 DOI: 10.1002/macp.200350065.

(52) Harrisson, S.; Ercole, F.; Muir, B. W. Living Spontaneous Gradient Copolymers of Acrylic Acid and Styrene: One-Pot Synthesis of pH-Responsive Amphiphiles. *Polym. Chem.* **2010**, *1* (3), 326–332 DOI: 10.1039/B9PY00301K.

(53) Wang, S.; Poehlein, G. W. Investigation of the Sequence Distribution of Bulk and Emulsion Styrene–acrylic Acid Copolymers by 1H- and 13C-NMR. *J. Appl. Polym. Sci.* **1993**, *49* (6), 991–1001 DOI: 10.1002/app.1993.070490605.

(54) Herbert, I. R. Statistical Analysis of Copolymer Sequence Distribution. In *NMR spectroscopy of polymers*; Springer Netherlands: Dordrecht, 1993; pp 50–79.

(55) Lomonosov Moscow State University Supercomputing Center http://hpc.msu.ru.

**For Table of contents use only**

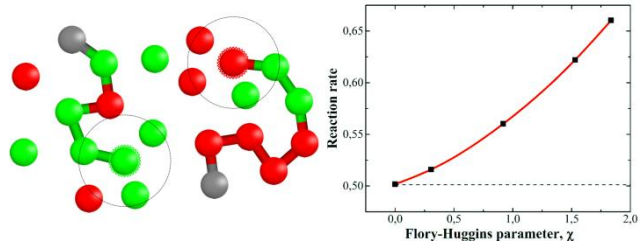

**Copolymerization of partly incompatible monomers: an insight from computer simulations.**

Alexey Gavrilov, Alexander Chertovich